\documentclass{an}
\usepackage{graphicx}
\usepackage{times}
\usepackage{fancyhdr}
\begin{document}

\title{X-ray Spectral States and High-Frequency QPOs in Black Hole Binaries}

\author{Ronald A. Remillard}
\institute{MIT Kavli Institute for Astrophysics and Space Research
        Massachusetts Institute of Technology, Cambridge MA 02139}

\date{Received; accepted; published online}

\abstract{
The high-frequency quasi-periodic oscillations (HFQPOs) 
in the X-ray emission of black hole binaries and candidates 
are briefly reviewed with respect to the question of X-ray states. 
We use the prescription of McClintock and Remillard (2005), 
in which X-ray spectra and power density spectra are used
to differentiate three main states of active accretion: "thermal", 
"hard", and "steep power-law". Respectively, each state emphasizes 
a different physical element that may contribute radiation: 
the accretion disk, a jet, and a compact non-thermal corona.  
We focus on the four sources that exhibit pairs of HFQPOs
with frequencies that scale in a 3:2 ratio.  All of these 
HFQPOs are associated with the steep power-law (SPL) state.  
Furthermore, there is a clear trend in which the $2 \nu_0$
oscillation is seen when the source has high X-ray luminosity,
while the $3 \nu_0$ QPO is seen at lower luminosity.}

\correspondence{rr@space.mit.edu}
\maketitle

%==========================================================
\section{X-ray States of Black Hole Binaries}  %===1==
\label{sec1}
%==========================================================
The X-ray states of black hole binaries have been defined
(McClintock and Remillard 2005) in terms of quantitative criteria 
that utilize both X-ray energy spectra and power density spectra (PDS).
This effort follows the lessons of extensive monitoring campaigns 
with the {\it Rossi} X-ray Timing Explorer ({\it RXTE}), which reveal 
the complexities of X-ray outbursts in black-hole binary systems 
and candidates (e.g. Sobczak et al. 2000; Homan et al. 2001).
These definitions of X-ray states utilize four criteria: $f_{disk}$,
the ratio of the disk flux to the total flux (both unabsorbed) at 2-20
keV; the power-law photon index ($\Gamma$) at energies below any break
or cutoff; the integrated rms power ($r$) in the PDS at 0.1--10 Hz,
expressed as a fraction of the average source count rate; and the
integrated rms amplitude ($a$) of a quasi-periodic oscillation (QPO)
detected in the range of 0.1--30 Hz. PDS criteria ($a$ and $r$)
are evaluates in a broad energy range, e.g. the full bandwidth of 
the {\it RXTE} PCA instrument, which is effectively 2--30 keV.

The energy spectra of accreting black holes often exhibit composite spectra 
consisting of two broadband components.  There is a multi-temperature 
accretion disk (Makishima et al. 1986 ; Li et al. 2005) and a power-law 
component (Zdziarski and Gi{\'e}rlinski 2004).  The thermal state designation 
selects observations in which the spectrum is dominated by the heat 
from the inner accretion disk.  The thermal state (formerly the 
``high/soft'' state) is defined by the following three conditions: 
$f > 0.75$; there are no QPOs with $a > 0.005$; $r < 0.06$.

There are two types of non-thermal spectra (Grove et al. 1998), and
they are primarily distinguished by the value $\Gamma$. There is a
hard state with $\Gamma \sim 1.7$, usually with an exponential
decrease beyond $\sim 100$ keV. The {\bf hard state} is associated
with a steady type of radio jet (Gallo et al. 2003; Fender 2005). In
terms of X-ray properties, the hard state is also defined with three
conditions: $f < 0.2$; $1.5 < \Gamma < 2.1$; $r > 0.1$. In the hard
state, the accretion disk spectrum may be absent, or it may appear to
be usually cool and large.

The other non-thermal state is associated with strong emission from a 
steep power-law component ($\Gamma \sim 2.5$), with no apparent cutoff
(Grove et al. 1998). This component tends to dominate black-hole binary 
spectra when the luminosity approaches the Eddington limit. 
Thermal emission from the disk remains visible during the SPL state. 
Low-frequency QPOs (LFQPOs), typically in the range 0.1--20 Hz, 
are frequently seen when the flux from the power law increases to the 
point that $f < 0.8$  The SPL state (formerly the ``very high'' state) 
is defined by: (1) $\Gamma > 2.4$, (2) $r < 0.15$, and (3) either $f < 0.8$,
while an LFQPO is present with $a > 0.01$, or $f < 0.5$ with no LFQPOs.

The temporal evolution of X-ray states for GRO~J1655--40 (1996-1997 
outburst), XTE~J1550--564 (1998-1999 outburst), and GX339-4 (several 
outbursts) are illustrated by Remillard (2005). Two of these examples 
display the opposite extremes of the complexity in black-hole outbursts.
GRO~J1655--40 shows a simple pattern of spectral evolution in which the 
thermal and SPL states evolve in proportion to luminosity, while 
XTE~J1550--564 shows complex behavior and intermediate states, in which 
there is a range of luminosity that is occupied by all states. 
This is interpreted as strong evidence that the primary variables
for understanding the energetics of accretion must include variables
in addition to the black hole mass and the mass accretion rate.

%==========================================================
\section{High-Frequency QPOs from Black Hole Binaries}  %===2==
\label{sec2}
%==========================================================

High-frequency QPOs (40-450 Hz) have been detected thus far in
7 black-hole binaries or candidates (see McClintock and Remillard 2005
and references therein).  These are transient and subtle
oscillations, with 0.5\% $< a < 5$\%. The energy dependence of
$a$ is more like the power-law spectrum than the thermal spectrum,
and some of the QPOs are only detected with significance in
hard energy bands (e.g. 6-30 keV or 13-30 keV).
For statistical reasons, some HFQPO detections additionally 
require efforts to group observations with similar spectral and/or
timing characteristics. 

Four sources (GRO~J1655-40, XTE~J1550-564, GRS~1915+105, and
H1743-322) exhibit pairs of QPOs that have commensurate frequencies in
a 3:2 ratio (Remillard et al. 2002; Homan et al. 2005; Remillard et
al. 2005; McClintock et al. 2005).  All of these HFQPOs have
frequencies above 100 Hz.  The observations associated with a
particular QPO may vary in X-ray luminosity by factors (max/min) of 3
to 8.  This supports the conclusion that HFQPO frequency systems are a
stable signature of the accreting black hole.  This is an important
difference from the the kHz QPOs in neutron-star systems, which 
show changes
in frequency when the luminosity changes.  Finally, for the three (of
four) cases where black hole mass measurements are available, the
frequencies of HFQPO pairs are consistent with a $M^{-1}$ dependence
(McClintock and Remillard 2005; $\nu_0 = 931 M^{-1}$).  This result is
generally consistent with oscillations that originate from effects of
GR, with an additional requirement for
similar values in the dimensionless BH spin parameter.
Thus, black hole HFQPOs with 3:2 frequency ratio may provide an
invaluable means to constrain black hole mass and spin via GR theory.

Commensurate HFQPO frequencies can be seen as a signature of an
oscillation driven by some type of resonance condition.  Abramowicz and
Kluzniak (2001) had proposed that QPOs could represent a resonance in the
coordinate frequencies given by GR for motions around a black hole
under strong gravity.  Earlier work had used GR coordinate frequencies
and associated beat frequencies to explain QPOs with variable
frequencies in both neutron-star and black-hole systems (Stella et
al. 1999), but without a resonance condition.

Current considerations of resonance concepts include more realistic
models which are discussed in detail elsewhere in these proceedings.
The ``parametric resonance'' concept (Klu{\'z}niak et al. 2004;
T\"or\"ok et al. 2004) describes oscillations rooted in fluid flow
where there is coupling between the radial and polar GR frequencies.
There is also a resonance model tied to asymmetric structures (e.g a
spiral wave) in the inner disk (Kato 2005).  Another alternative is to
consider that state changes might thicken the inner disk into a torus,
where the normal modes under GR (with or without a resonance
condition) can yield oscillations with a 3:2 frequency ratio (Rezzolla
et al. 2003; Fragile 2005).  Finally, one recent MHD simulation
reports evidence for resonant oscillations (Kato 2004).  This research
will be studied vigorously, while more than one model
might be relevant for the different types of QPOs
in accreting BH and NS systems.

\begin{figure*}
\resizebox{\hsize}{!} 
{\includegraphics[angle=-90, width=\hsize]{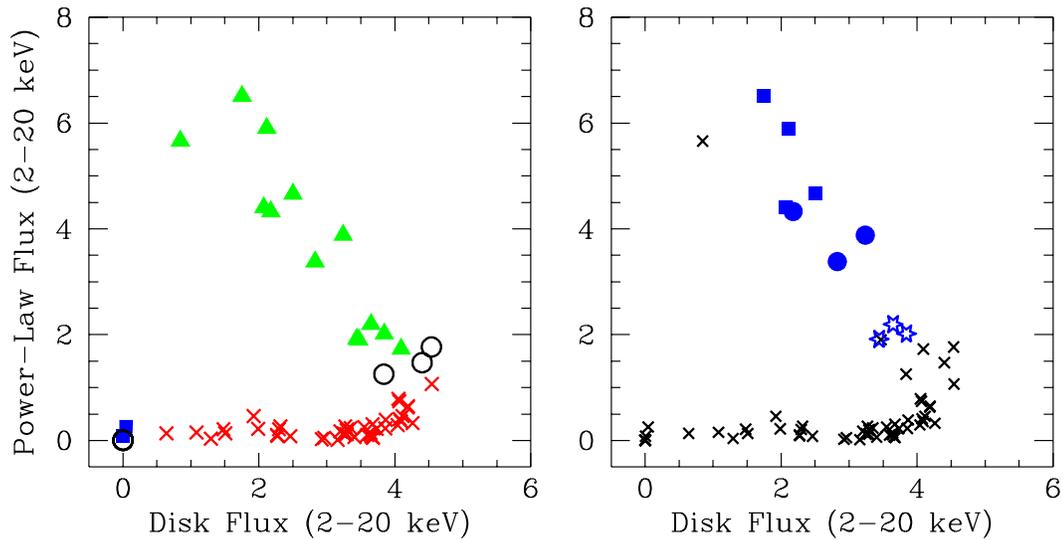}}
\caption{ X-ray states and HFQPOs during the 1996-1997 outburst of
GRO~J1655--40. The left panel shows the energy diagram, where flux
from the accretion disk is plotted versus flux from the power-law
component.  Here, the symbol type denotes the X-ray state: thermal
(red ``x''), hard (blue square), steep power-law (green triangle), and
any type of intermediate state (yellow circle). The right panel shows
the same data points, while the symbol choice denotes HFQPO
detections: 300 Hz (blue squares), 450 Hz (blue star), both HFQPOs
(blue circle), and no HFQPO (black ``x'').  The HFQPO detections
are clearly linked to the SPL state, and the HFQPO frequency is
clearly correlated with power-law luminosity. }
\label{fig1} 
\end{figure*}

%==========================================================
\section{High-Frequency QPOs and the SPL State}  %===3==
\label{sec2}
%==========================================================

A study of the HFQPOs in GRO~J1655-40 and XTE~J1550-564 has shown
that detections in individual observations are associated with the SPL
state (Remillard et al. 2002). In Figs. 1 and 2, this point is made
more clearly by comparing HFQPO detections to X-ray state
classifications that utilize the criteria of McClintock and Remillard
(2005).  Each plot displays an energy diagram, where the flux from the
accretion disk is plotted versus the flux from the power-law
component.  The flux is determined from the parameters obtained from
spectral fits.  Here the flux is integrated over the range of 2--20
keV, which is the band used to define X-ray states.  It has been
shown that the results can also be displayed in terms of bolometric
fluxes without changing the conclusions (Remillard et al. 2002).

In the left panels of Figs. 1 and 2, the X-ray state of each
observation is noted via the choice of the plotting symbol. The state
codes are: thermal (red x), hard (blue square), SPL (green triangle),
and any intermediate type (yellow circle). The 1996-1997 outburst of
GRO~J1655-40 (Fig. 1, left) is mostly confined to the softer
X-ray states (i.e. thermal and SPL).  On the other hand, observations
of XTE~J1550-564 (Fig. 2, left) show far greater complexity, with
a mixture of states for a wide range in luminosity.
These data combine the 1998-1999 and 2000
outbursts of the source, since HFQPOs are seen during both outbursts.
The determinations of fluxes follows the same procedures 
described for GRO~J1655-40.

The right panels of Figs. 1 and 2 show the same data points, but the
choice of symbols is related to the properties of HFQPOs.
Observations without HFQPO detections are shown with a black
``x''.  HFQPO detections are distinguished for frequency: $2 \nu_0$
oscillations (blue squares), $3 \nu_0$ oscillations (blue star).  For
GRO~J1655-40 (only), there are three observations that show both $2
\nu_0$ and $3 \nu_0$) HFQPOs, and the data are shown
with filled blue circles.  Comparisons of the left and right panels of
Figs. 1 and 2 show that HFQPO detections for the two sources are all
obtained during the SPL state.  These figures also
display the clear association of $2 \nu_0$ HFQPOs with higher
luminosity in the power-law component, while $3 \nu_0$ HFQPOs occur at
lower luminosity, as has been reported previously (Remillard et
al. 2002).

The HFQPO detections with a 3:2 frequency ratio for the other two BH
sources require more complicated data selections, and so we cannot
compare X-ray states and HFQPO properties in the same manner.  In the
case of H1743-322, there are very few HFQPO detections in individual
observations (Homan et al. 2005; Remillard et al. 2005).  However, the
technique used to combine observations to gain some of the HFQPO
detections at $2 \nu_0$ and $3 \nu_0$ utilized SPL classifications
grouped by luminosity level.  The success of this strategy shows that
H1743-322 follows the same patterns displayed in Figs. 1 and 2.  In
GRS1915+105, the HFQPO pair with 3:2 frequency ratio involves
extraction of oscillations from portions of two different modes of
unstable light curves with cyclic patterns of variability.  Further
analysis is required to produce the energy diagrams for these data.

\begin{figure*}
\resizebox{\hsize}{!} 
{\includegraphics[angle=-90, width=\hsize]{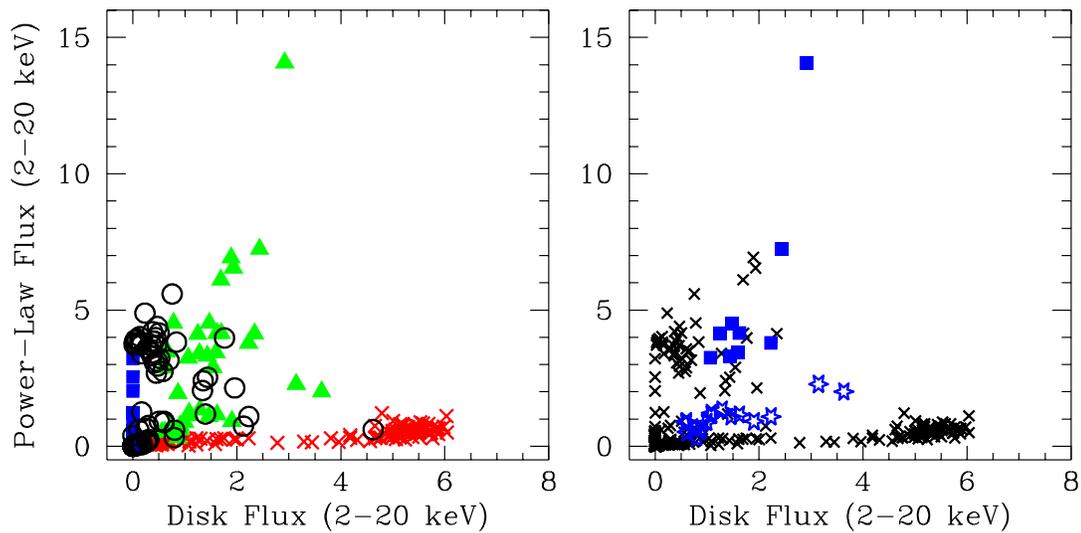}}
\caption{ X-ray states and HFQPOs during the 1998-1999 and 2000
outbursts (combined) of XTE~J1550--564. The left panel shows the
energy diagram, with plotting symbol chosen to denote the X-ray state,
as in Fig. 1.  The right panel shows frequency-coded HFQPO detections:
near 184 Hz (blue squares), near 276 Hz (blue star), and no HFQPO
(black ``x'').  Again, HFQPO detections are linked to the SPL state,
and the HFQPO frequency is correlated with power-law luminosity. }
\label{fig1} 
\end{figure*}

There are undoubtedly statistical issues to consider in judging the
absence of HFQPOs in Figs. 1 and 2, since most detections are near the
detection limit (i.e. 3--5 $\sigma$).  Nevertheless, efforts to group
observations in the hard and thermal states, in order to lower the
detection threshold, routinely yield null results at either the known
or random HFQPO frequencies.  We conclude that the models for
explaining HFQPO frequencies must also explain the geometry,
energetics, and radiation mechanisms for the SPL state.

%%%%%%%%%%%%%%%%%%%%%%%%%%%%%%%%%%%%%%%%%%%%%%%%%%%%%%%%%%%%%%%%%%%%%%

%==========================================================
\section{Spectral states, QPO, modulation of X-rays}      %===5==
%==========================================================

%\acknowledgements

\end{document}